# Programming at Exascale: Challenges and Innovations

Jalal Abdulbaqi

Department of Electrical and Computer Engineering, Rutgers University, New Brunswick, NJ, USA

jalal.nazar@rutgers.edu

ABSTRACT

Supercomputers become faster as hardware and software technologies continue to evolve. Current supercomputers are capable of $10^{15}$ floating point operations per second (FLOPS) that called Petascale system. The High Performance Computer (HPC) community is Looking forward to the system with capability of $10^{18}$ (FLOPS) that is called Exascale. Having a system to thousand times faster than the previous one produces challenges to the high performance computer (HPC) community. These challenges require innovation in software and hardware. In this paper, the challenges posed for programming at Exascale systems are reviewed and the developments in the main programming models and systems are surveyed.

## I. INTRODUCTION

Exascale computing refers to computing with systems that deliver performance in the range of $10^{18}$ (exa) floating point operations per second (FLOPS) [1]. The high performance computers (HPC) available today have the ability to do no more than 33 peta ($10^{15}$) FLOPS [2]. This increase in computational ability can enhance the development and discoveries in many engineering and science fields such as climate modeling, protein folding, drug discovery, energy research and data analysis [3]. Resizing of the new systems can change the whole architecture of the HPCs. Therefore, a massive change is required in the programming models and systems. This change can be implemented either by evolving them or by introducing new programming technology.

A parallel programming models and systems are continue evolve to exploit the development in the HPC hardware. A programming model is different than to programming system (language, library, API, etc.). A programming model is a set of software technologies that express parallel algorithms and match applications with the underlying parallel systems, while a programming system is an implementation of one or more of these models. Since 2003, parallel programming development became a very important issue because the increase of the microprocessor frequency became unpractical due to heat dissipation and energy consumption issues [4]. Parallel programming technology that available today are still not enough to utilize the current hardware as well as the new Exascale systems, which require programming roles, such as the control of data movement.

The programming for Exascale systems faces several challenges required to addressed. Power, concurrency, memory, communication, resiliency, and heterogeneity are the major challenging aspects that encounter the implementation of Exascale systems [5],[6],[7],[8],[9]. Developing programming models and systems for Exascale machines can face issues related to the challenges mentioned [6]. In this paper, the challenges that pose the programmers at Exascale systems are reviewed and the developments in the main programming models and systems are surveyed.

In the next section, a review of the challenges is presented from the perspective. In section III, a survey for the developed in the programming models and systems is illustrated. In section IV, hybrid programming systems are explained. The conclusion gives a summary and discusses works that are still under development.

## II. CHALLENGES

Developing programmers for Exascale systems pose challenges more than what we have for current systems. These challenges are the result of the vast change in hardware architecture that required to implement Exascale systems. Since 2008, many reports have addressed these challenges to achieve Exascale [5], [6], [10]. According to these reports, the main challenges are power, concurrency, memory, communication, resilience, and heterogeneity.

The increase in the consuming power cannot be the same as that for computational capability. In fact, the desired objective of consuming power to reach Exascale should not exceed 20 MW [7]. This limit has an effect on of both hardware and software implementation. Most of the consumed power is either in data movement [6] or chip temperature [8]. Data movement includes the communication to access the memory or the inter-process communication [6]. Therefore, new power-aware algorithms are needed [7] that can reduce the movement of data [6] and do the load balancing [8]. Both of these solutions require an explicit user control of communication [8].

The increase in computational ability comes from the increase in cores on a chip and support threads. This increment in the concurrency is at least 1000x than the current systems [11]. The bulk-synchronous execution model commonly used today is not able to scale to this new limit. New models and algorithms have to be developed to limit the synchronization and communication [7]. The nested bulk-synchronous model, which is an extension to the normal bulk-synchronous model and dynamic nested parallelism, can be used as an evolved model [8]. Another way is by scheduling parallel tasks dynamically, but this will make the debugging process harder [7].

As a consequence of the memory technology that we have currently, memory size is not expected to increase as well as processing elements. In addition, while the memory in the each node will increase, the memory for each core will decrease [7]. The power consumed by the current memory technology is relatively high. Many new kinds of memory are expected to be used in the Exascale systems as well as nonvolatile memory, stacked memory, scratchpad memory, processor-in-memory, and deep cache hierarchies [7]. In addition, it is expected that the memory hierarchy will be deeper [6]. As a result of these changes, a new memory algorithms-aware are required and more explicit management are essential [6], [7].

There are two main kinds of data transfer in HPC systems: memory bandwidth and nodes communications. Both play a bottleneck role in the overall system speed. Communication also is not expected to increase as the same way like the number of the processing units [7]. The increment in the data movement can affect not only the system performance but also the power consumed. In fact, most of the expected power in the Exascale system will be consumed in the communication [6]. Data locality is the main issue that should be considered in designing the new algorithms to minimize the data movement [7].

The tremendous increase in the hardware failure that results from a large number of Exascale system components is one of the most challenges in such system. When an error occurs in the current system, it is either corrected automatically or it will cause an exception which will stop the process. The current technique to deal with this exception is using global checkpoint and restart. Unfortunately, such global recovery can no longer handle the increased number of errors. The key solution is using local recovery besides taking the advantage of the algorithms in the application. Conclusively, several development efforts address the resilience issues for Exascale system can be found in [12].

Increasing the node concurrency and saving the consume power can be achieve by using the accelerators. However, using accelerators such as GPU and coprocessor can increase the programming model complexity. Simplicity and performance are the tread off benchmark used for developing programming models and systems. Other developing efforts work on utilizing the computational power of the accelerators over the distributed system, which increase the complexity and involve the communication issue [13]. In fact, different types of models can be combined (hybrid programming) to achieve this goal.

Finally, the extensive assumption can be a serious challenge for Exascale systems. Regularly, varying of system features and realization can affect the productivity and consistency of applications [9]. Therefore, the models have to maintain the portability.

## III. PROGRAMMING MODELS AND SYSTEMS

Several parallel programming models and systems are developed to utilize the multi-core processors, many-core accelerators, and distributed systems. A programming model is "an abstraction of the underlying computer system that allows for the expression of both algorithms and data structures" [9]. While, a programming system is an implementation of one or more models, (e.g. programming language, library, framework, API, etc). Programming models can be categorized according to the memory topology that is shared memory, distributed memory, and distributed shared memory as shown in Fig. 1. In distributed systems, another categorized had according to communication protocols that is two-sided communication and one-sided communication. One-sided communication also called Remote Memory Access (RMA) which is "a single process calls a function, which updates either local memory with a value from another process or remote memory with a value from the calling process." [3]. In following subsection, more detailed mentioned about the main three category of the programming models and the associate programming systems.

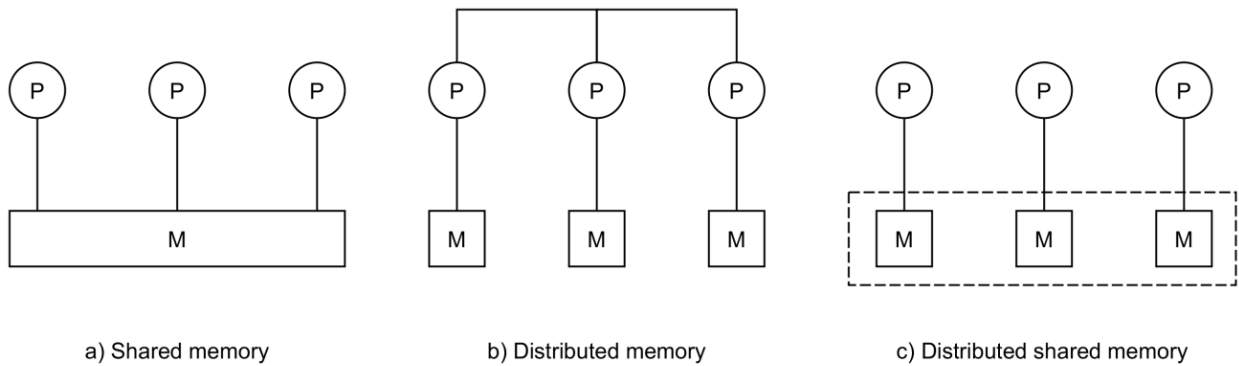

a) Shared memory      b) Distributed memory      c) Distributed shared memory

Fig. 1 Memory Topologies (Models).
Memory are represented as squires, while processes are represented with circles.

*A. Multithreading*

The multithreading programming model is based on the shared memory model, in which the executing program components (processes/threads) share a single common address. The multithreading programming model is useful not only for parallel programming but also to enhance the performance of some serial programming issues such as interacting with slow I/O system or with a slow network connection [14].

The earlier implementation of this model is as a set of C programming language extension called the Portable Operating System Interface (POSIX) or Pthreads. Pthreads implemented as a library that creates and manipulate the threads activities explicitly. Pthreads is a low level, an unstructured programming system that used by experts to develop specific applications, but it is not useful for general large-scale problems.

The second important implementation of multithreading model is Open Multi-Processing (OpenMP). OpenMP is an application programming interface (API) implemented as a library and compiler directives. These directives (#pragmas) used to create threads, perform synchronization, and manage the memory. OpenMP is the most used shared memory programming model in the current HPC's.

OpenMP 4.x brings new features such as supporting the accelerators and thread affinity. the first feature is essential for the Exascale system as mentioned in section II while the last provide the user more control on threads execution, which is useful to improve the locality [15]. Therefore, OpenMP is expected to continue play an important role in Exascale era either as a single system or as a hybrid system with may be MPI, more information in section IV. On the other hand, other efforts try to extending the OpenMP to distributed memory model such as Jay Hoeflinger in [16] who introduce this idea, which leads to the Cluster OpenMP by Intel®, but the project has discontinued in 2010 [17]. The evaluation research shows unacceptable performance for OpenMP as a distributed memory model [18], [19].

*B. Message Passing*

The message passing programming model is based on the distributed memory model, where the processes communicate among them by sending and receiving messages. Portability and explicit control over the process memory location are the two main features of message passing model [14]. The message passing model implementation called Massage Passing Interface (MPI). MPI is "a message-passing library interface specification" [20]. The Exascale system architecture can be as current Petascale system that is a distributed memory model. Therefore, MPI is expected to mimic an important role in Exascale systems because it is the most popular one for distributed memory parallelism. MPI can be used either as a single programming system or combined with others. MPI tested successfully to work on over 1 million cores, which give us a positive sign to its scalability to work in Exascale systems [8]. MPI is under the research focus for years ago to investigates the challenges that pose its scalability and other issues at extreme scale conditions.

Thakur, Rajeev and et.al. in [21] have reviewed the challenges posed the MPI, which mainly the scalability that affected by performance and memory consumption. Scalability issue is the most challenge that faced by both MPI specification and implementation. In MPI specification, four issues have suffered from scalability, some functions argument size, graph topology, all-to-all communication, and the representation of process ranks. Most of these issues are fixed in MPI-3 version and if not you can use another deferent algorithm such as all-to-all communication and process ranks scalability problem. Resilience and low performance one-sided communication are another challenges pose MPI which also addressed in MPI-3. In MPI Implementation, scalability is still the dominating problem. Process mappings, the creation of new communicators, MPI Init, and collective communication are experienced a scalability issue. The solution implemented in MPICH2 for process mapping and suggestions are presented for the creation of new communications, but no clear solution to MPI startup using Init. Finally, Using hybrid algorithms will help to deal with scalable issues in algorithms for collective communication.

Gropp and Snir in [8] address the challenges that pose the programming technologies at Exascale systems. Again, scalability is the main issue that affect MPI. The amount of message buffer space, the description of the MPI communicators, and the MPI processes start in the number of processes can be linear (non-scalable). The first one can be resolved by proper implementation while a trade off between time and memory space can be helpful for MPI communicators. Although, scalable start-up systems already exist for MPI processes start. The all to all communication scalability issue is determined by introducing all to some communication pattern that is scalable. Furthermore, the bulk-synchronous, which supported by MPI, has extreme concurrency and asynchronous. MPI resolve this problem by implement another synchronous methods. Finally, the general resolution for MPI at Exascale should be focused on scalability implementation either by the MPI developers or programmers.

Da Costa, Georges, et al. in [13] include three limitations for MPI to work at Exascale. The first limitation is that MPI has a static data distribution, which not suitable to cope with dynamic load balancing. The second one is that the non-scalable all-to-all communication. Finally, the limitation of collective access to the I/O request and the data partitioning.

*C. Partitioned Global Address Space (PGAS)*

Partitioned Global Address Space (PGAS) is a programming model suited for shared and distributed memory parallel machines [22]. this model is implemented according to the distributed shared memory model. This implementation distinguishes between the local data, which located inside the node and the remote data, which located inside other nodes. PGAS goals are productivity and performance. Productivity comes from implementing the global shared address space. High performance and scalability can be achieved by distinguishing between the local and remote data access [23].

In the late 1990s, three PGAS languages introduced, Co-Array Fortran (CAF), Titanium, and Unified Parallel C (UPC). These languages are an extensions to the programming languages, Fortran, Java, and C respectively. The logic behind CAF is that accessing remote data is controlled explicitly [23]. The CAF extension is implemented in several compilers such as Cray, g95, GNU Fortran, Intel Fortran Compiler, Rice (CAF 2.0), and OpenUH. Titanium is dialect of Java developed at UC Berkeley to support scientific computing on HPC. Titanium is introduced to work on massively parallel supercomputers and distributed systems with targets, safety, portability, and support for building complex data structures. UPC provides capability for direct user specification of program parallelism and control of data distribution and access [24]. UPC support both bulk-synchronous and fine-grain parallel programming models [25].

In the early 2000s, the Defense Advanced Research Projects Agency (DARPA) create the High Productivity Computing Systems (HPCS) program to develop a new programming languages taking in consider the performance, portability, reliability, and programmability [26]. The HPCS produce three programming languages, Chapel, X10, and Fortress. Chapel, the Cascade High Productivity Language is parallel programming language developed by Cray®. Chapel support any parallel algorithm on any hardware architecture with a special optimization for Cray machines. Chapel supports definite concepts for describing parallelism from locality. Chapel is developed to support higher or lower programming levels, as required by the programmer. Chapel also involves a set of sequential language aspects such as type inference, iterator functions, object-oriented programming, and a rich set of array types. X10 is a parallel object-oriented programming language developed by IBM. X10 extents the PGAS model to asynchronous PGAS with two core concepts: places and asynchrony. Places are the collection of cells partitioned memory as chunks. An async spawns a new thread that operates asynchronously with other threads. Fortress is a discontinued programming language developed by Sun Microsystems® then it is owned by Oracle®. Fortress features included implicit parallelism, Unicode support and concrete syntax similar to mathematical notation.

*D. Heterogeneous Programming*

Heterogeneous Programming refers to the programming the system that has accelerator as well as the CPU cores. Accelerator can be Graphical Processing Unit (GPU), Accelerated Processing Unit (APU), many-core coprocessor, etc. The heterogeneous programming introduced in early 2000's when NVIDIA® announces their first programming GPU [27]. Accelerators are manufactured by several companies such as NVIDIA programmable GPU series (such as Tesla, Fermi, Kepler, and Maxwell), AMD Fusion series (such as Trinity, Kaveri, and Carrizo), and Intel Xeon Phi coprocessor series (such as Knights Corner, Knights Landing, and Knights Hill).

The heterogeneous architecture that include CPU and accelerators require new programming models and systems. Compute Unified Device Architecture (CUDA) was developed by NVIDIA in 2006 to use with Tesla GPU for non-graphics application programming. CUDA is a programming platform and application programming interface (API) designed to work with C/C++ and FORTRAN programming languages.

Open Computing Language (OpenCL) is a framework developed originally by Apple®, but later it is introduced by Khronos Group, a non-profit technology consortium. OpenCL is an industry standard not only to work with CPU's and GPU's but also to work with another type of processing units such as digital signal processors (DSPs), field-programmable gate arrays (FPGAs) and other hardware accelerators. OpenCL includes a parallel programming language and API, which used to select the hardware that works with [25].

Cilk plus and Threading Building Blocks (TBB) are other tools maintained by Intel® to program and utilize the Xeon Phi coprocessor as well as Intel processors. Cilk plus is a multithreaded general-purpose programming language while TBB is a C++ parallel programming library for multi-core processors.

Open Accelerators (OpenACC) is a programming standard developed by Cray, CAPS, NVIDIA and PGI designed to simplify the heterogeneous programming. Inspired by OpenMP,

OpenACC uses compiler directives with C/C++ and FORTRAN programming languages to accelerated their source code.

C++ Accelerated Massive Parallelism (C++ AMP) is a compiler and programming model extension to C++ programming language developed by Microsoft® that enables the acceleration of C++ code on data-parallel hardware such as GPU [28]. C++AMP is an open specification and a DirectX 11 implementation, which intent to performance portability by permitting the same code to run either on a GPU or a CPU [29].

Exascale implementation requires a very large number of cores inside each node, as mentioned in section II. Heterogeneous architecture can fulfill this challenge as consequence of the large number of cores included in the accelerators. The use of the accelerators in HPC had been increased due to the enhancement in performance and power efficiency [13]. Heterogamous programming faces two main challenges, software cost and performance portability. software cost means the complexity of the development and maintaining applications while the performance portability refers to the ability of the application to work on different devices without losing its efficiency [30].

OpenCL support wide range of CPU, GPU, and FPGA devices, but it has poor cache performance on CPU. The main drawback in OpenCL is the software cost, which is the difficulty to make changes in its code. OpenACC developed to make the code changing easy by using the compiler directives for loops that inspired by OpenMP. Open MP 4.0 also evolved to support accelerators, which facilitate the programming and maintaining the code. Although, OpenMP shows a low performance when it compared with other GPU programming systems such as CUDA and OpenACC [31], [32]. However, both OpenCL and OpenACC are not adaptable to the classic programming language such as C. So that, it is still harder to debug the code than for C programming. Therefore, C++AMP built to address this problem by automatically parallelize the C++ loops and handling the data movement using the compiler. On the other hand, all of the mentioned programming tools still having problems in achieving the performance portability for different kinds of computations. Li-Wen Chang and et. al. introduce a new programming system called TANGRAM [33], [34], which is a general-purpose high-level language that achieves high performance for heterogamous architectures. TANGRAM evaluated on different applications using CPU, and two types of NVIDIA GPUs and the performance is distinctive with respect to CUDA 7 and OpenMP 4.0.

## IV. HYBRID PROGRAMMING

Hybrid programming is combining two or more of the programming systems mentioned in section III. The common combination is between shared-memory and distributed memory programming systems. This combination called MPI+X, where X refers to the multithreading programming model implementation. The aim of hybrid programming is to have the capability of both systems e.g. the simplicity of shared-memory and the scalability of the distributed-memory systems [27]. According to The International Exascale Software Roadmap state that "Hybrid Programming is a practical way to program Exascale systems" [35]. As mention in section II, Exascale required to have a deep hierarchical memory and communication. Therefore, programming models also need to develop or combine in a hierarchical way. Although that hybrid programming is seen as the solution for the future complicated systems, there are several restrictions result from the combining programming systems. The main issue arise from hybrid model is that both combined the programming systems can consume the recourse such as memory and communication more that each system aside [8]. This limitation need to be addressed because we already know that the memory in the Exascale system will not increase linearly like the computational ability.

## V. CONCLUSION

Programming at Exascale era required a big change in evolving and innovating in parallel programming models and systems. OpenMP and MPI are the most used and evolved programming systems in the current HPCs for shared and distributed memory models respectively. It is expected for these systems to continue used at Exascale systems as a shared memory (OpenMP), distributed -memory (MPI), or both (MPI+OpenMP) [36].

PGAS languages expected to be more important at Exascale because of the distinct features and the development efforts. PGAS languages can be implemented either as a single system or even for a hybrid programming with MPI. Despite the fact that PGAS has several advantages for future HPC, it has two disadvantages make it not suitable for non-HPC systems. First, PGAS assume that all processes work on similar hardware architecture. Second, PGAS model not support the dynamically spawning multiple activates [27]. However, compaction studies between MPI and some PGAS languages showed a distinct performance of MPI over PGAS for certain applications.

Finally, programming models and system still need to evolve and harmonize between different types to work together in an optimized hybrid system that can face the high scalability and deep hierarchy architecture.

[12]  F. Cappello, A. Geist, and W. Gropp, "Toward exascale resilience: 2014 update," *Supercomput. ...*, 2014.

[13]  G. Da Costa, T. Fahringer, J.-A. Rico-Gallego, I. Grasso, A. Hristov, H. D. Karatza, A. Lastovetsky, F. Marozzo, D. Petcu, and G. L. Stavrinides, "Exascale Machines Require New Programming Paradigms and Runtimes," *Supercomput. Front. Innov.*, 2015.

[14]  J. Dongarra, I. Foster, G. Fox, and W. Gropp, *Sourcebook of parallel computing*. 2003.

[15]  OpenMP Architecture Review Board, "OpenMP Application Programming Interface," no. November. 2015.

[16]  J. Hoeflinger, "Extending OpenMP to clusters," *White Pap. Intel Corp.*, 2006.

[17]  J. Hoeflinger, "Cluster OpenMP* for Intel® Compilers." [Online]. Available: https://software.intel.com/en-us/articles/cluster-openmp-for-intel-compilers.

[18]  C. Terboven and D. A. Mey, "First experiences with intel cluster openmp," *OpenMP a New Era ...*, 2008.

[19]  J. Cai and A. Rendell, "Predicting Performance of Intel Cluster OpenMP with Code Analysis Method," *ANU Comput. Sci. ...*, 2008.

[20]  Message Passing Interface Forum, "MPI: A Message-Passing Interface Standard Version 3.1," 2015.

[21]  R. Thakur, P. Balaji, and D. Buntinas, "MPI at Exascale," *Procceedings ...*, 2010.

[22]  G. Almasi, "Encyclopedia of Parallel Computing," D. Padua, Ed. Boston, MA: Springer US, 2011, pp. 1539–1545.

[23]  M. Wael, S. Marr, and B. Fraine, "Partitioned global address space languages," *ACM Comput. ...*, 2015.

[24]  T. El-Ghazawi, W. Carlson, T. Sterling, and K. Yelick, *UPC: distributed shared memory programming*. 2005.

[25]  P. Balaji, *Programming models for parallel computing*. MIT Press, 2015.

[26]  J. Dongarra, R. Graybill, W. Harrod, and R. Lucas, "DARPA's HPCS program: History, models, tools, languages," *Adv. ...*, 2008.
16